\documentclass[apj]{emulateapj}
\pdfoutput=1
\usepackage{mathptmx}
\usepackage{ulem}

\def\gtorder{\mathrel{\raise.3ex\hbox{$>$}\mkern-14mu
             \lower0.6ex\hbox{$\sim$}}}
\def\ltorder{\mathrel{\raise.3ex\hbox{$<$}\mkern-14mu
             \lower0.6ex\hbox{$\sim$}}}

\slugcomment{The Astrophysical Journal, 797:71 (4pp), 2014 December 10}
\shorttitle{Probing the IGM with FRBs}
\shortauthors{Zheng et al.}

\begin{document}

\title{Probing the Intergalactic Medium with Fast Radio Bursts}
\author{
Z. Zheng\altaffilmark{1}, 
E. O. Ofek\altaffilmark{2},
S. R. Kulkarni\altaffilmark{3},
J. D. Neill\altaffilmark{4}, 
and
M. Juric\altaffilmark{5} 
}
\altaffiltext{1}{Department of Physics \&\ Astronomy, University of Utah,
115 South 1400 East \#201, Salt Lake City, UT~84112, USA}
\altaffiltext{2}{Department of Particle Physics \&\ Astrophysics, 
Weizmann Institute of Science, Rehovot 76100, Israel}
\altaffiltext{3}{Caltech Optical Observatories 249-17,
California Institute of Technology, Pasadena, CA~91125, USA}
\altaffiltext{4}{Space Radiation Laboratory 290-17,
California Institute of Technology, Pasadena, CA~91125, USA}
\altaffiltext{5}{Department of Astronomy, University of Washington, Box 351580,
	Seattle, WA~98195, USA}

\begin{abstract}
The recently discovered fast radio bursts (FRBs), presumably of extra-galactic
origin, have the potential to become a powerful probe of the intergalactic 
medium (IGM). We point out a few such potential applications.
We provide expressions for the dispersion measure and rotation measure as a 
function of redshift, and we discuss the sensitivity of these measures to 
the \ion{He}{2} reionization and the IGM magnetic field. Finally we calculate 
the microlensing effect from an isolate, extragalctic stellar-mass compact 
object on the FRB spectrum. The time delays between the two lensing 
images will induce constructive and destructive interference, leaving a 
specific imprint on the spectra of FRBs. With a high all-sky rate, a large 
statistical sample of FRBs is expected to make these applications feasible.
\end{abstract}


\section{Introduction}
\label{sec:Introduction}

Variability of cosmological radio sources has long been proposed to probe 
the properties of inter-galactic medium (IGM). \citet{hs65} suggested to 
detect IGM dispersion measure (DM) through the variability of the radio signal 
from quasars, aiming at using the inferred DM to distinguish different 
cosmological models. \citet{Weinberg72} and \citet{Ginzburg73} suggested the 
use of radio flares to measure the DM and thus probe the IGM density.  Later, 
radio emission from gamma-ray bursts (GRBs) and their afterglows were proposed 
as means to determine distances to GRBs and to probe the IGM \citep{Palmer93}, 
to study the prehistory of GRBs \citep{lpl97}, and to constrain the hydrogen
re-ionization history of the universe (\citealt{Ioka03,Inoue04}).
However, radio quasars and GRBs, despite the aspirations of the authors,
simply lack sharp features that allow their signals to be easily used to 
probe the intervening electrons in the IGM. 

The situation can completely
change with the recently discovered short radio bursts. The first such burst
was reported by \citet{lbm+07}, which is an intense (30\, Jy) and short 
duration (5-ms) burst at 1.4\, GHz (named as the {\it Sparker} in 
\citealt{KONZJ14}). Following the above discovery, \citet{tsb+13} reported 
the finding of four short duration bursts with an estimated all-sky rate of
$10^4$ events day$^{-1}$ (denoted as ``Fast Radio Bursts'' or FRBs). 
\citet{Spitler14} discovered one FRB in the Arecibo Pulsar ALFA Survey.
These 
short radio bursts show DMs of order of a few hundred 
to thousand, suggesting a substantial contribution from electrons in the 
IGM. \citet{KONZJ14} performed a thorough investigation of the {\it Sparker}
and FRBs to explore possible constraints on sites or processes to explain
such high DMs, and concluded that they are of extra-galactic origin, provided 
that the inferred DM arises due to propagation through cold plasma. A variety
of models have been proposed for the progenitors of such short bursts 
\citep[e.g.,][]{pp07,Vachaspati08,fr14,Totani2013,Zhang2014,lhr+14,kim13,
KONZJ14}.  

The pulse nature, the high rate, and the extra-galactic origin make the
{\it Sparker}-like events and FRBs well suited for being used as a potentially
powerful probe to the IGM. Since the discovery of the short duration bursts, 
DM measurements have been proposed to probe missing baryons around halos of 
galaxies \citep{McQuinn14}, study the baryon content in the IGM \citep{dz14}, 
and constrain cosmology and the equation of state of dark energy \citep{gao14,zlw+14}.

In this paper, we explore further potential applications of a 
{\it Sparker}-like population or FRBs in probing the IGM (for simplicity, 
hereafter we call the {\it Sparker}-like events and FRBs collectively as FRBs).
Specifically, we first point out the use of them to probe the 
era of \ion{He}{2} reionization and intergalactic magnetic field 
(\S\ref{sec:ProbeIGM}) and then comment about the potential use of FRBs for 
detecting a cosmological population of massive compact halo objects, 
i.e., MACHOs (\S\ref{sec:IntergalacticMachos}). Finally, we give a summary in
\S\ref{sec:summary}.

\section{Probing \ion{He}{2} Reionization and IGM Magnetic Field}
\label{sec:ProbeIGM}

In this section, we specifically focus on the potential use of FRBs
to probe the era of \ion{He}{2} re-ionization and IGM magnetic field.  

\ion{He}{2} reionization can be regarded as the last phase transition in the 
universe, after the major one related to Hydrogen and \ion{He}{1} reionization 
above $z\sim 6$. Stars are not hot
enough to ionize \ion{He}{2} with an ionization potential of 54.4\,eV.
However, there is some evidence that \ion{He}{2} re-ionization occurred at
$z\sim 3$ from the transmission of the \ion{He}{2} Ly$\alpha$ forest and
the temperature change of the IGM [see Furlaneto \&\ Oh (2007a,
b)\nocite{fo07a}\nocite{fo07b} and references therein].  The ionization can
be caused by soft X-ray emission from activate galactic nuclei (AGNs) and
hard ionizing photons from quasars. Thus,
observations of \ion{He}{2} re-ionization provide a new diagnostic of the
build-up of the AGN and quasar population.  The same observations also provide
important clues to the structure and thermal evolution of the IGM, which is
related to the missing baryons problem \citep[e.g.,][]{Gnat11}.  Clearly
there is great value in using FRBs in our study of
cosmology and AGNs/quasars.

\begin{figure}
\centerline{\includegraphics[width=8.5cm]{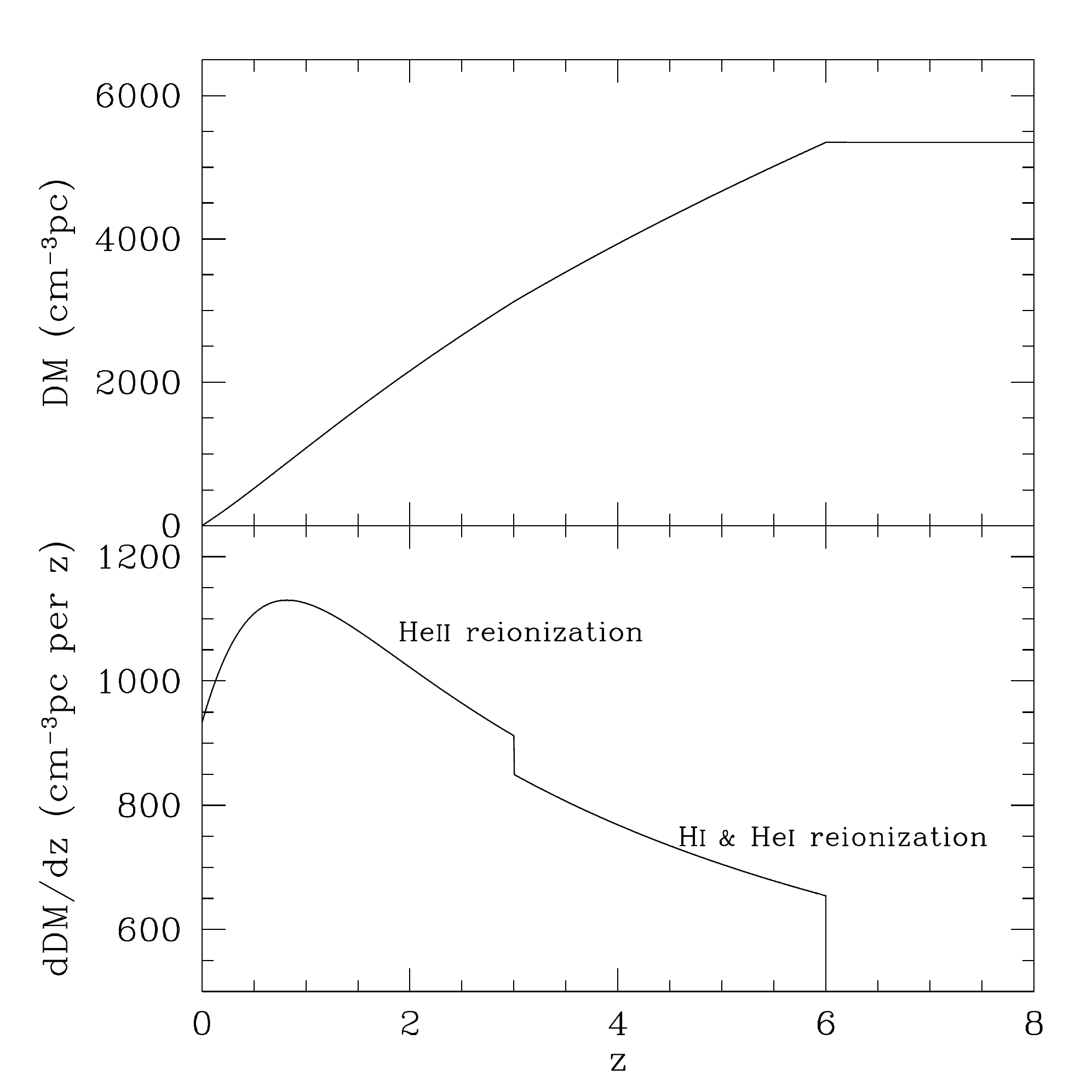}}
\caption{Illustration of using Dispersion Measure to probe the epoch
of re-ionization of \ion{He}{2}.  Top and bottom panels show
DM and its derivative as a function of redshift, respectively. A
sharp \ion{H}{1} and \ion{He}{1} re-ionization at $z\sim 6$ and a
sharp \ion{He}{2} re-ionization at $z\sim 3$ are assumed.
\label{fig:DMz}} 
\end{figure}

For the purpose of computing the DM from the IGM, there are three effects on 
the propagation time, $t_p$, of a photon traveling through the IGM to reach the
observer from a cosmological distance: the continuous change of the
photon's frequency, $\omega$, due to the redshift of light, the change of
the plasma frequency, $\omega_p^2=4\pi n_e(z)e^2/m_e$, due to the change in
the IGM electron density, $n_e(z)$, with redshift, and the time dilation
effect. The first two effects lead to a change in the group velocity,
$v_g=c(1-\omega_p^2/\omega^2)^{1/2}$, with redshift. The propagation time
of a photon emitted at redshift $z$ seen by an observer at redshift 0 is
then
\begin{eqnarray}
	\label{eqn:tp}
	t_p&=&\int_0^z dz\frac{dl}{dz} \frac{1}{v_g} (1+z), \cr
	   &=&\int_0^z \frac{c dz}{(1+z)H(z)} 
	\frac{1}{c}\left(1+\frac{1}{2}\frac{\omega_p^2}{\omega^2}\right)(1+z),
\end{eqnarray}
where $H(z)=H_0[\Omega_m(1+z)^3+\Omega_\Lambda]^{1/2}$ is the Hubble constant 
at $z$, with $\Omega_m$ the matter density parameter and 
$\Omega_\Lambda=1-\Omega_m$ (assuming a spatially flat universe), and the last
$(1+z)$ factor accounts for time dilation. The frequency, $\omega$,
is related to the observed frequency, $\omega_{\rm obs}$, through
$\omega=(1+z)\omega_{\rm obs}$.

The IGM electron density $n_e(z)$ can be expressed as
\begin{eqnarray}
	n_e(z) &=& n_0(1+z)^3\left[(1-Y)f_{\rm HII}+\frac{1}{4}Y(f_{\rm
	HeII}+2f_{\rm HeIII})\right],\cr
	&=& n_0(1+z)^3f_e(z),
\end{eqnarray}
where
\begin{equation}
	n_0 =\frac{\Omega_b\rho_c}{m_H}=2.475\times 10^{-7}
	\left(\frac{\Omega_b h^2}{0.022}\right) {\rm cm}^{-3}
\end{equation}
is the mean number density of nucleons at $z=0$. Here $\Omega_b$ is the
baryon density in units of the $z=0$ critical density $\rho_c$ and $h$ is 
the $z=0$ Hubble constant in units of 100 ${\rm km\, s^{-1}\, Mpc^{-1}}$.
Since we assume to observe a large number of FRBs at each redshift, 
it is appropriate to use the mean density of the IGM in the calculation 
without worrying about the density fluctuations.
In the expression,
$Y\simeq 0.25$ is the mass fraction of helium, $f_{\rm HII}$ is the
ionization fraction of hydrogen, and $f_{\rm HeII}$ and $f_{\rm
HeIII}$ are the ionization fractions of singly and double ionized
helium. After helium re-ionization ($z\sim 2-3$), we essentially
have $f_{\rm HII}=1$, $f_{\rm HeII}=0$, and $f_{\rm HeIII}=1$, which
gives $f_e\simeq0.88$ at low redshifts.

The observed dispersion measure (DM)  is defined as
\begin{equation}
	\frac{dt_p}{d\omega_{\rm obs}}
	=-\frac{4\pi e^2}{cm_e\omega_{\rm obs}^3} {\rm DM}.
\end{equation}
In combination with equation~(\ref{eqn:tp}), we have
\begin{eqnarray}
	\label{eqn:DM}
	{\rm DM}& = & n_0\frac{c}{H_0}
	\int_0^z\frac{dz (1+z) f_e(z)}{\sqrt{\Omega_m(1+z)^3+\Omega_\Lambda}} 
	\nonumber, \\
	& = & 1060\, {\rm cm}^{-3}{\rm pc} 
	\left(\frac{\Omega_b h^2}{0.022}\right)
	\left(\frac{h}{0.7}\right)^{-1}\cr
	& & \times \int_0^z\frac{dz (1+z) f_e(z)}{\sqrt{\Omega_m(1+z)^3+\Omega_\Lambda}}.
\end{eqnarray}
For a constant $f_e$, the above integral can be approximated as
\begin{eqnarray}
	{\rm DM} & \cong & 933\, {\rm cm}^{-3}{\rm pc} 
         \left(\frac{f_e}{0.88}\right)
         \left(\frac{\Omega_b h^2}{0.022}\right)
         \left(\frac{h}{0.7}\right)^{-1} \cr
        & & \times
         \Bigg[ \left(\frac{\Omega_m}{0.25}\right)^{0.1} a_1 (x-1) 
         \ + 
	 \left(\frac{\Omega_m}{0.25}\right) a_2 (x^{2.5}-1) \cr
        & & 
         \,\,\,\,\,\, + \left(\frac{\Omega_m}{0.25}\right)^{1.5} a_3 (x^4-1)
         \Bigg],
\end{eqnarray}
with $x=(1+z)^2$, $a_1=0.5372$, $a_2=-0.0189$, and $a_3=0.00052$.  The
accuracy of this approximation is better than $\sim$2\% for $z<1.5$.
At low redshifts, one can use the following approximation,
\begin{eqnarray}
{\rm DM} &\cong& 933\ {\rm cm^{-3}\ pc}\Big[z+(0.5-0.75\Omega_m)z^2\Big] \cr
  & & \times \left(\frac{f_e}{0.88}\right) \left(\frac{\Omega_b h^2}{0.022}\right) \left(\frac{h}{0.7}\right)^{-1},
\end{eqnarray}
which has a 5\% accuracy up to $z=0.6$. For a constant $f_e$, the integral in
equation~(\ref{eqn:DM}) shares some similarity with the expression of the 
luminosity distance $D_L$, with the $(1+z)$ factor pulled out of the integral
in the latter. In terms of $D_L$, the integral can be approximated as
\begin{equation}
{\rm DM} \cong n_0 f_e D_L \left[1+0.932z+(0.16\Omega_m-0.078)z^2\right]^{-0.5},
\end{equation}
which has an accuracy $\lesssim$0.5\% for $0<z<3$ with $0.25<\Omega_m<0.35$.

As an illustration, the DM as a function of $z$ is displayed in 
Figure~\ref{fig:DMz}.  This is an idealized plot since we assume a sharp  
\ion{He}{2} re-ionization at $z\sim
3$. The re-ionization is better seen in the slope or derivative of the DM
curve. The jump is about 8\%.  Whether this jump will be seen or not will
depend very strongly on the contribution to the DM of FRBs
by the electrons in the host galaxies and whether FRBs
can be found to redshifts as high as $z\sim 3 $.

Similarly, we can obtain the rotation measure (${\rm RM}$). The Faraday
rotation is
\begin{equation}
	\Delta\theta = \frac{2\pi e^3}{m_e^2 c^2\omega_{\rm obs}^2} n_0B_0\frac{c}{H_0}
	\int_0^z\frac{f_e(z)b_\parallel(z) dz}{\sqrt{\Omega_m(1+z)^3+\Omega_\Lambda}},
\end{equation}
where $b_\parallel(z)\equiv B_\parallel(z)/B_0$ is the line-of-sight
magnetic field, $B_\parallel(z)$, in units of the local IGM magnetic
field, $B_0$.  We then have
\begin{eqnarray}
	{\rm RM} &=& 8.61 \,{\rm rad}\,{\rm m}^{-2}\left(\frac{\Omega_b h^2}{0.022}\right) 
	 \left(\frac{h}{0.7}\right)^{-1} \left(\frac{B_0}{\rm 10nG}\right) \cr
        & & \times
         \int_0^z\frac{f_e(z)b_\parallel(z) dz}{\sqrt{\Omega_m(1+z)^3+\Omega_\Lambda}}.
\end{eqnarray}
At low redshifts, where we approximate $f_e=0.88$ and $b_\parallel=1$,
the {\rm RM} can be written as 
\begin{eqnarray} 
\label{eqn:RM}
{\rm RM} &\cong&
	7.57(z-0.75\Omega_m z^2)\,{\rm rad}\,{\rm m}^{-2} \cr
        & & \times
	\left(\frac{f_e}{0.88}\right)
         \left(\frac{\Omega_b h^2}{0.022}\right)
         \left(\frac{h}{0.7}\right)^{-1} \left(\frac{B_0}{\rm 10nG}\right).
\end{eqnarray}

So far, there are no accurate measurements for the magnetic field in the
IGM with densities of the order of the mean density (see \citealt{kbm+07}).
We note that the local IGM magnetic field would have a strength of
4($T_{\rm IGM}/10^4K$)nG if energy equipartition is assumed.
Radio-synchrotron radiation has been detected in the Coma super-cluster
\citep{kkg+89}, implying a field strength of 0.3 to 0.6 $\mu$G. RM
measurements of FRBs can constraint the magnitude of 
IGM magnetic field and its evolution, providing clues on its
origin.

In addition, if the DM and RM have a large contribution from a scattering 
screen, measurement of RM will provide a strong clue to the location of the
scattering (and thus dispersion) screen. If the scattering arises in the
IGM, then the RM from the IGM is less than 30\,rad\,m$^{-2}$ for $z\ltorder
0.3$ \citep{kbm+07}. A much larger RM can be produced if the screen is
located in the host galaxy.

\section{Probing Intergalactic MACHOs}
\label{sec:IntergalacticMachos}

Another potential use of FRBs is to constrain the
existence of floating MACHO-like objects in the IGM via gravitational
lensing (e.g., \citealt{Gould92, spg93, mnn+99}).  

A fortunate
alignment of an intervening point mass object of mass $M$ with
a FRB will result in two images.  The time delay for each
image, with respect to a ray that arrives by the shortest path, is
the sum of a geometric term and a gravitational delay term and is
given by \citet{nb96} as
\begin{equation}
 	\label{Eq:timedelay}
 	t(\theta)=\frac{1+z_l}{c}\frac{4GM}{c^2}
	\left(\frac{1}{2}\frac{\theta^2}{\theta_{\rm E}^2}-\ln|\theta|\right),
\end{equation}
where $z_l$ is the redshift of the lens, $\theta$ is the angular
distance between the lens and the image, and $\theta_{\rm E}$ is
the annular radius of the Einstein ring,
\begin{equation}
	\label{Eq:ThetaE}
	\theta_{\rm E} = \sqrt{\frac{4GM}{c^{2}} 
	\frac{D_{ls}}{D_lD_s}}.
\end{equation}
Here, $D_l$, $D_s$, and $D_ls$, are the observer-lens,
observer-source, and lens-source angular diameter distances,
respectively.  The positions of the two images are
\begin{equation}
	\label{Eq:Theta}
	\theta_\pm=\frac{1}{2}\left(b\pm\sqrt{b^2+4\theta_{\rm E}^2}\right),
\end{equation}
where $b$ is the impact parameter (i.e., source-lens angular distance).

From equations~(\ref{Eq:timedelay}) and (\ref{Eq:Theta}), the lensing 
time delay between the two images is
\begin{eqnarray}
	\Delta t_l&\equiv& t(\theta_-)-t(\theta_+)\cr
	&=&\frac{1+z_l}{c}\frac{4GM}{c^2}
	\left[\frac{1}{2}u\sqrt{u^2+4}
	+\ln\left(\frac{\sqrt{u^2+4}+u}{\sqrt{u^2+4}-u}\right)\right],
\end{eqnarray}
where $u\equiv b/\theta_{\rm E}$ is the impact parameter in units of the
Einstein ring radius. For a typical value of $u=1$:
\begin{equation}
	\Delta t_l=41(1+z_l)(M/1M_\odot)\mu{\rm s}.
\end{equation}

The short time-scales of FRBs make the effect of relative motions between
the source, lens, and observer completely negligible (e.g., the fractional 
change in $u$ is of the order of $10^{-12}$ during a lensed milli-second FRB 
event at cosmological distances with relative motion of 100 
${\rm km\, s^{-1}}$). So unlike the case for Galactic MACHOs, we do not 
expect to observe microlensing light curves for MACHO-like objects in the IGM.
However, the two images have different amplifications,
$A_\pm=1/2\pm(u^2+2)/(2u\sqrt{u^2+4})$ and the time delay ensures
they also have different phases. Therefore, the two images undergo
constructive and destructive interference. The total amplification is
\begin{equation}
	\label{Eq:amplification}
	A(\omega)=\frac{u^2+2}{u\sqrt{u^2+4}}+
	\frac{2}{u\sqrt{u^2+4}}\cos(\omega\Delta t_l),
\end{equation}
where $\nu=\omega/(2\pi)$ is the frequency in the observer's frame.  The
spectrum of the lensed FRB will thus consist of maxima and minima with the 
separation of two
maxima (or minima) being $\Delta\nu=\Delta t_l^{-1}$, which is tens of kHz
for $M\sim 1M_\odot$ and $u\sim 1$ -- well within the reach of current
technology. Note that $\Delta\nu$ is independent of the observing
frequency, if the time delay is purely caused by lensing.  However, in
general $\Delta t_l$ in equation~(\ref{Eq:amplification}) should be
replaced by the sum of all possible time delays. There could be dispersive
delay due to the rays suffering different DM or delay due to multi-path
propagation from a scattering screen.  For FRBs, this latter effect destroys the
coherence\footnote{Equivalently, the de-coherence is
the size of the scattered disk exceeding the image separation with the
attendant loss of visibility function.} of the rays.  For this reason,
observing at higher frequencies (i.e., $\gtorder5$\,GHz) is highly
desirable.  Furthermore, a matched-filter
approach\footnote{naturally obtained via the
cross-correlation function in a XF-type interferometer.} can additionally
improve the detection.

Finally, we can estimate the optical depth for lensing (e.g.,
\citealt{nb96}). For a proper number density $n(z)$ of lenses with mass $M$ in
a spatially flat universe, the optical depth for lensing to sources at 
redshift $z_s$ is
\begin{eqnarray}
\tau(z_s) & = & \int_0^{z_s} n(z) \pi (D_l\theta_E)^2 dD_c/(1+z)\cr
          & = & \frac{4\pi GM}{c^2 D_s}\int_0^{z_s} n(z) D_{ls} D_l dD_c/(1+z),
\end{eqnarray}
where the angular diameter distances are $D_s=D_A(0,z_s)$, 
$D_l=D_A(0,z)$, $D_{ls}=D_A(z,z_s)$, and 
$D_A(z_1,z_2)=\int_{z_1}^{z_2} dD_c/(1+z_2)$, with $dD_c=cdz/H(z)$ 
the comoving distance element. In the case of proper number density evolving
as $n_0(1+z)^2$ [i.e., comoving number density evolving as $n_0/(1+z)$], the 
optical depth can be put in a form similar to the result in the static 
Euclidean space,
\begin{equation}
\label{eqn:tau}
\tau = \frac{2\pi}{3} \frac{G\rho_{l,0}}{c^{2}} D_c^{2} 
     \simeq 0.014\, \Omega_l (D_c/ 1 {\rm Gpc})^{2},
\label{Eq:LensingTau}
\end{equation}
where $\rho_{l,0}=n_0M$ is the mass density of lenses at $z=0$ and $\Omega_l$ 
is this mass density in units of the $z=0$ critical density of the universe. 
In the case of a constant comoving number density, the result is about 44\% 
higher than that in equation~(\ref{eqn:tau}) for $z_s=1$.

\section{Summary and Discussion}
\label{sec:summary}

The pulse nature, the high rate, and the extra-galactic origin make FRBs
ideal for probing the IGM. In this paper, we present potential 
applications of FRBs to probe \ion{He}{2} reionization, IGM magnetic field,
and MACHO-like objects in the IGM.

For these applications, a large population of FRBs are necessary. The 
\ion{He}{2} reionization causes a small change in the DM. It leads to
a $\sim 8$\% jump in the differential DM across the \ion{He}{2} reionization 
epoch. At least a few hundred FRBs around $z\sim 2-3$ is needed to detect such
a change. Our Galaxy, FRB host galaxies, and any intervening galaxies or 
clouds with free electrons can add scatter in the DM. Clearly, more FRBs and
searching for host galaxies are desired to constrain and understand such a 
scatter. It is likely that we need thousands of FRBs around $z\sim 2-3$ to 
learn about the \ion{He}{2} reionization from the DM measurements. 

The probe of IGM magnetic field with FRBs could be more challenging, given 
that its strength is likely of the order of nG while that inside a galaxy is
of the order of $\mu$G. The RM from a typical galaxy is about 812 rad m$^{-2}$
$(n_e/{\rm cm}^{-3})(B_\parallel/\mu G)(l/{\rm kpc})$ (with $l$ the path 
length). The RM from IGM to $z=1$ is only 6 rad m$^{-2}$ for an IGM magnetic
field of 10 nG according to equation~(\ref{eqn:RM}). If the scatter in 
the galaxy-caused RM is of the same order of $\sim 800$ rad m$^{-2}$, tens 
of thousand of FRBs are needed to clearly map out the redshift evolution of
RM caused by the IGM magnetic field.

For the MACHO-like objects in the IGM, the upper bound for their density 
parameter $\Omega_l$ is $\Omega_m$. If they are of baryonic origin (e.g.,
stellar remnants), which may be more likely, the upper bound is then 
$\Omega_b$. According to equation~(\ref{eqn:tau}), we expect the lensing 
optical depth to be (much) less than $6\times 10^{-4}$ to a distance of 1 Gpc,
which requires at least tens of thousand FRBs to discover the events with 
the lensing signal we point out.

The current estimated all-sky rate of FRBs is $10^4$ events day$^{-1}$ 
\citep{tsb+13}. Therefore, with well-designed and dedicated FRB surveys, 
all the above requirements of a large statistical sample of FRBs are not 
demanding at all. Such surveys would provide an invaluable opportunity to 
advance our understanding of the IGM.

\acknowledgments

We thank Shude Mao for useful comments.
Z.Z. was partially supported by NSF grant AST-1208891 and NASA grant NNX14AC89G.
M.J. acknowledges the support of the Washington Research Foundation through 
its Data Science Chair and the University of Washington Provost's Initiative 
in Data-Intensive Discovery.
E.O.O. is incumbent of the Arye Dissentshik career development chair
and is grateful to support by grants from the Willner Family
Leadership Institute Ilan Gluzman (Secaucus NJ), Israeli Ministry
of Science, Israel Science Foundation, Minerva and the I-CORE Program
of the Planning and Budgeting Committee, and The Israel Science
Foundation. S.R.K. thanks the hospitality of the Institute
for Advanced Study (IAS). The sylvan surroundings and verdant
intellectual ambiance of IAS resulted in a fecund mini-sabbatical
stay (Fall 2007).  

\bibliographystyle{apj1b}
\bibliography{bibsparker}
\end{document}